A NEW CLASS OF SETI BEACONS THAT CONTAIN INFORMATION

G. R. Harp, R. F. Ackermann, Samantha K. Blair, J. Arbunich, P. R. Backus,

J. C. Tarter, and the ATA Team

SETI Institute

515 N. Whisman Road

Mountain View, CA  94043

www.seti.org

1      Abstract.

In the cm-wavelength range, an extraterrestrial electromagnetic narrow band (sine wave) beacon is an excellent choice to get alien attention across interstellar distances because 1) it is not strongly affected by interstellar / interplanetary dispersion or scattering, and 2) searching for narrowband signals is computationally efficient (scales as $N_S \log(N_S)$ where $N_S$ = number of voltage samples). At the SETI Institute we use the Allen Telescope Array (ATA) to simultaneously search for continuous and pulsed sine wave signals, hence the term "narrowband," (Backus, 2010). We assume the alien transmitter is always on, or if they are sending narrow band pulses, the time between pulses is at least 3 times smaller than the time of a SETI observation on one point of the sky and one observing frequency (typically 100 seconds). We shall return to these assumptions below.

Unfortunately, a *continuous* narrowband beacon carries only one bit of information (that the alien civilization exists) and carry zero symbol rate after the beacon is discovered. In contrast, modest to wide bandwidth signals can support modest to large symbol rates (information rates). This advantage is offset because any wideband signal suffers corruption during its voyage from transmitter to receiver, caused by the Interstellar medium or ISM. The ISM disperses and scatters the signal, resulting in e.g. pulse broadening. Most wideband signals require additional computation such as for a de-dispersion search, the start and end time, the frequency bandwidth, etc. making such searches hopelessly inefficient with current technology. Historically, wideband SETI searches in the cm-wavelength range signals have hardly been pursued. A special case search for broadband pulses (Siemion, 2010) for SETI was recently performed. These searches mitigate the computational problem with very fast FPGA hardware, but the task is still difficult.

Here we consider a special case wideband signal where two or more delayed copies of the same signal are transmitted over the same frequency and bandwidth, with the result that ISM dispersion *and* scattering[1] cancel out during the detection stage. Such a signal is both a good beacon (easy to find) and carries arbitrarily large information rate (limited only by the atmospheric transparency to ~10 GHz). The discovery process uses an autocorrelation algorithm, and we outline a compute scheme where the beacon discovery search can be accomplished with only 2x the processing of a conventional sine wave search, and discuss signal to background response for sighting the beacon. Once the beacon is discovered, the focus turns to information extraction. Information extraction requires similar processing as for generic wideband signal searches, but since we have already identified the beacon, the efficiency of information extraction is negligible.

2     Introduction.

When designing a signal for interstellar transmission to an unknown but technologically competent species, we must consider how that species (e.g. humans) might discover the signal <u>as</u>

---

[1] In general, arbitrary scattering destroys information, however this method allows beacon discovery even in the presence of strong scattering.

distinct from the galactic background radiation. Note that SETI is different from most earth-based communication problems because 1) we don't know, *a priori*, how much uninteresting naturally-generated power is arriving from any random point on the sky and 2) we must invent a process that separates the SETI signal from the background noise (which includes both the galactic background and the receiver noise.

At the same time, most SETI researchers suspect that an extra-solar civilization will wish to communicate nontrivial information to humans. Until now, the primary focus of radio SETI observations has been narrowband signals or strong broadband pulses. These signals can be used only for "beacons," since they convey no "message" beyond a single symbol of information. It is usually suggested that the message information will be communicated in an entirely different signal mode (or with extremely low symbol rate consistent with the narrowband criterion).

We introduce a new signal type where the beacon and message are encoded in one and the same signal. The proof of principle described here opens the road for invention of more SETI beacons that have this property. The point of this paper is to show that a signal that is both easily discovered and contains a message is possible and takes no more than a factor of 2 times the computational power of a narrowband SETI search. Another advantage is that our proposal searches an "orthogonal" space to narrowband or pulsed SETI, opening an uncharted territory for exploration. One disadvantage of the proposed signal type is that in with the algorithms presented here, it gives lower signal power to background power ratio than narrowband SETI for equal power transmitters.

We propose to recover signals using autocorrelation spectroscopy. While a more generalized concept was first suggested in 1965 (Drake, 1965), to the best of our knowledge no prior published work on radio SETI searches using autocorrelation spectroscopy exists. We begin this paper with an extended introduction to put this work into perspective, followed by a description of the technique, and finishing with preliminary observations taken with the Allen Telescope Array (ATA) that prove the concept of the technique and detection algorithm.

2.1     Some Constraints Due to the Galactic Background Radiation.

Starting with the extragalactic background, it is a natural law of the universe that almost all galactic radiation arises from sources with relatively large bandwidths (between 500 Hz masers (Cohen, 1987) and $10^{19}$ Hz gamma ray bursts). In the radio frequency range natural signals have time-varying amplitude which in a narrow bandwidth that is not distinguishable from Gaussian white noise (Figure 1). This is true even in the case of narrow spectral lines or masers; in the frequencies of emission the signals are noise-like. In the time domain, there are pulsars with time variations as small as 1 ms (Backer, 1982). But these unusual sources have recognizable characteristics that allow SETI researchers to identify them. For this reason we endeavor to exclude noise-like signals (in the absence of further information[2]) from our search space as being most probably associated with natural sources.

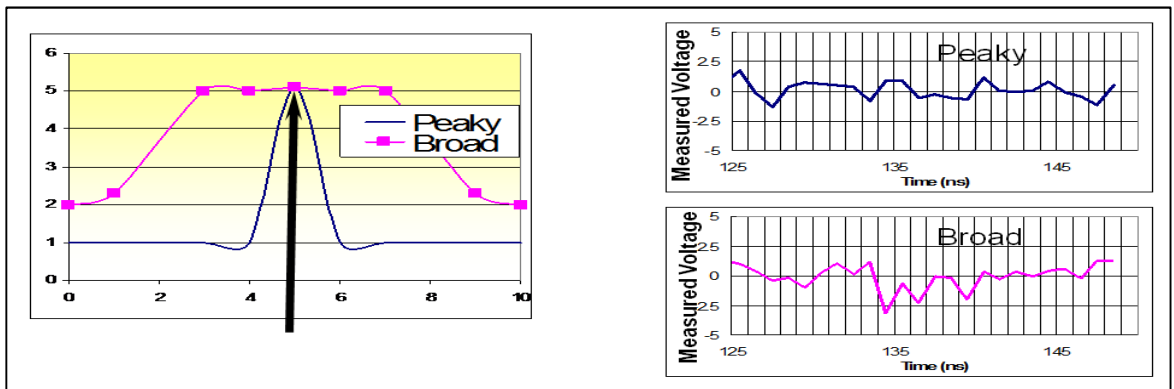

Figure 1: Left: Highly schematic representation of two fictitious radio signals represented in the frequency domain: A broadband source (Broad) and a relatively narrow band source such as spectral line radiation (Peaky). The gains of the two features are normalized and measured in a narrow

---

[2] Notice that some artificial signals may be "noise-like," i.e. spread spectrum signals. A spread-spectrum algorithm could distinguish these signals from real background radiation and noise.

bandwidth centered at the black arrow. The voltage signals from these sources might look like the two graphs on the right, which show Gaussian random noise for the electric field amplitude as it arrives at the telescope. This is an important quality of natural signals and successful SETI observations must focus on non-Gaussian random signals.

Most galactic radiation is indistinguishable from Gaussian random noise because it obtains from a large number of unresolved independent radiators situated very far from the telescope. For the radiation to be visible across vast distances, a great number of similarly radiating sources is required since those sources radiate incoherently and usually have separations greater than the light crossing time between radiators (though there are exceptional cases). According to Rice's mean value theorem, radiators may have any spectral character (or frequency occupation) and will tend to Gaussian random noise in a narrow bandwidth as the number of sources $\rightarrow \infty$. Such natural radiation is compounded with the receiver noise which is also nearly Gaussian white noise in character.

2.2    Information Content.

Nyquist's and Shannon's theorems (Nyquist, 1928; Shannon, 1949) give us some insight. For a given signal, the maximal received information rate (symbol rate) is roughly equal the bandwidth of the signal detector. For example, a continuous sine wave has zero symbol rate (symbols per second) if the signal appears in only one frequency bin of the detector.[3] Similarly, when singular wideband pulses arrive, we obtain only one bit of information. Comparatively, many wideband signal types allow information recovery with a symbol rate up to its received frequency bandwidth.

---

[3] Note that the signal might contain information discoverable by a different detector with e.g. smaller bin size. However, for a given detector, once the detection is made no signals encoded in a single detector bin may be recovered.

Generally, SETI transmitters can never take full advantage of the transmitted bandwidth because 1) a maximally compressed information signals are appear as Gaussian white noise hence is *indistinguishable* from natural radiation, and 2) a corollary to Shannon's theorem is that a maximal symbol rate is obtained when there is no redundancy between one sample and the next, making it impossible to decode (without the key). We can find/decode beacons only if they contain substantial redundancy, i.e. the signal must strike a balance between redundancy (which makes it noticeably artificial), and detectability / decodeability. Here we are assuming that we have no prior knowledge of the transmitted signal. If we know (by some means) or guess a part of the encoded information, then this requirement is relaxed. An example of this is narrowband pulsed SETI, where we "guess" that signal form is essentially a sine wave, but carries information in very slow (few Hz) pulses.

An example of extreme redundancy is the sine wave or single-pulse signal which contain only 1 bit of information but are easily noticeable against the galactic background.

2.3     Some Constraints Due to the Inter-Stellar Medium (ISM).

To reach us the signal must traverse the space between transmitter and receiver. This space is filled with dust, neutral gas and ionized gas. Between 1-10 GHz, the most important source of signal distortion is the free electrons in the ionized gas or plasma (mostly ionized hydrogen). While traversing plasma, electromagnetic (EM) photons acquire an effective rest mass equal to $\hbar\omega_p$, and travel more slowly than the speed of light. The plasma angular frequency $\omega_p$ is given by (Jackson, 1975).

$$\omega_p = \frac{4\pi\rho_e e^2}{m_e} \tag{1.1}$$

where $\rho_e$ is the free electron density (electrons per cubic meter). The constants $e$, and $m_e$ are the electron charge and electron mass, respectively. The average interstellar medium has about 1 electron per cubic centimeter (J. M. Cordes, Lazio, T. J. W., 1992; J. M. Cordes, Lazio, T. J. W., & Sagan, C., 1997) leading to plasma frequency $\omega_p \sim 0.1$ rad/s $\sim 0.16$ Hz $\sim 4 \times 10^{-17}$ eV. If the ISM plasma were homogenous with this value, EM frequencies below this cutoff would not propagate. Higher frequency EM waves would propagate, but more slowly than the speed of light, with higher frequencies travelling faster (Fitzpatrick, 2006). In the radio frequency range 1-10 GHz, for at transmitter located a distance L from the receiver, the light travel time $\tau$ for the signal is given by

$$\tau \cong \frac{L}{c}\left(1 + \frac{\omega_p^2}{2\omega^2}\right). \qquad (1.2)$$

Another derivation of this delay can be found in (Thompson, 2001) p. 576. With this expression it is straightforward to simulate signal distortion in a uniform ISM plasma. This is done in Figure 2 where a single pulse with length 50 nanoseconds (left) is dispersed according to Equation (1.2). After traversing only the short distance (4 LY) the pulse is been broadened by a factor of 40 and arrives at the detector with 30x lower peak power and reducing the signal to background radiation level by a factor of 6-30.

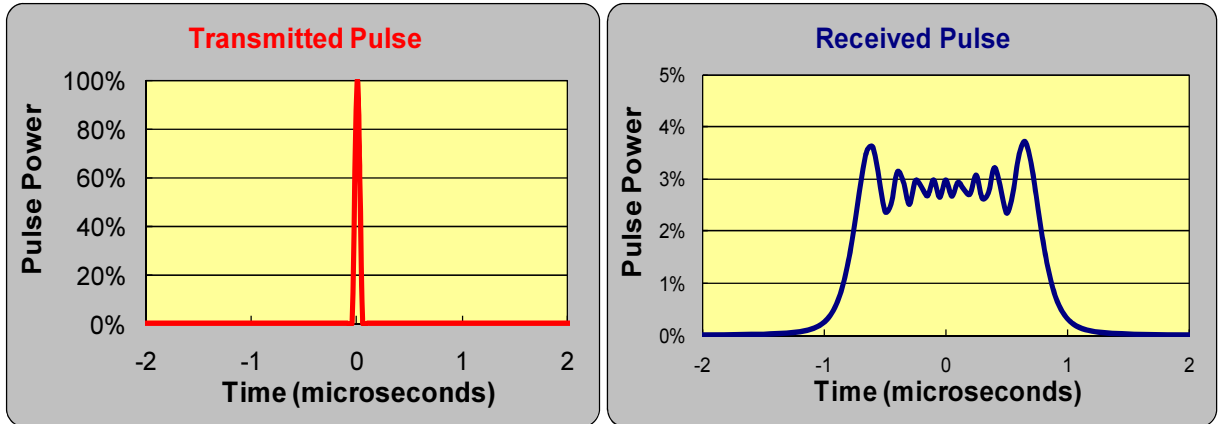

Figure 2: Left: A 50 ns pulse as observed near the transmitter located the same distance from Earth as the nearest star (Proxima Centauri). Right: The dispersed pulse as received by a detector on Earth. The received pulse is not quite symmetrical about Time = 0 because dispersion is in nonlinear in frequency.

First-order dispersion correction (de-dispersion) can be done even when the plasma is not homogeneous as long as the electron column density or "dispersion measure," $DM \equiv \int_0^L \rho_e \, dL$ between transmitter and receiver is known. De-dispersion recovers a narrow pulse from the dispersed pulse on the right hand side of Figure 2, and works equally well with other kinds of wideband signals. Unfortunately, the DM of received signals is generally not known since the interstellar medium is clumpy on all length scales and often we do not know the distance to the source. The result is that for recovery of any wideband signal, such as the pulse above, a search over a large number of test values for DM is required (unless you have prior knowledge, such as when the pulse was sent[4]), which makes searches for wideband signals computationally inefficient as compared with narrowband signals.

For any but the nearest transmitters, it is also necessary to understand the non-linear scattering of signals caused by the inhomogeneous electron density in the ISM. Scattering defects are variously known as pulse broadening, signal fading, and scintillation. Cordes et al. (J.

---

[4] Since a pulse search requires a search over both start time and DM, it is possible to order the searches by 1) performing a search over start times followed by an easy estimate of DM, or 2) by performing the DM search followed by an easy time of arrival estimate. It may be possible to optimize the computation in pulse searches by choosing the appropriate order.

M. Cordes, Lazio, T. J. W., & Sagan, C. , 1997; Lazio, 2002) conclude, "…scintillations are very likely to allow initial detections of narrowband signals, while making redetections extremely improbable... This conclusion holds for relatively distant sources but does not apply to radio SETI toward nearby stars (~100 pc)." The reason re-detections are improbable is that even continuous signals from distant sources will fade in and out at the receiver location (analogous to listening to a radio station at long-distance). Scintillation is negligibly small at short distances (< 100 pc) and lower frequencies (< 3 GHz). Scintillation grows monotonically larger at greater distances and higher frequencies. For example, at 1 GHz and distances > 500 pc, 100% fading (in and out) is expected. See references for more detail.

2.4     Conventional SETI and Pulse Searches.

Because of dispersion and scattering, conventional SETI searches have focused on continuous or slowly (few Hz) pulsed signals that are nearly monochromatic in a given reference frame. Such signals propagate through the ISM with little or no corruption (apart from fading) since a multipath sum of a monochromatic signal cannot change its frequency. At the SETI Institute, observers look for "drifting" signals with bandwidth of <10 Hz and Doppler drift rate up to 1 Hz / second. It is important to search over Doppler drift space because our Earth is accelerating as it rotates about its axis, the Sun, and galactic center. The same can be said for the transmitter.[5] Such acceleration causes compression / dilation of the signal in the time domain, leading to time-dependent frequency changes for the monochromatic signal. The Doppler drift of a narrowband signal is proportional to its frequency for a given relative acceleration. This variation is partially mitigated by the acceptance of signals with bandwidth up to 10 Hz. For the purposes of this paper, we shall not consider the frequency variation of drift rate and consider drift rates up to 1 Hz/s as being an acceptable range to allow most SETI transmitters to be observed at any frequency.

---

[5] This search can be obviated by the assumption of a "global" reference frame such as the one associated with the galactic center.

In "coherent" searches over short time periods (e.g. 1 second) Doppler drift is negligible and a simple Fourier Transform[6] can be applied to detect a narrowband signal. This detection algorithm is considered fast and scales as $N_S \ln(N_S)$ where $N_S$ is the number of samples in the measurement.

To its detriment, a search for pulses like that in Figure 2 would require the same searches over frequency (and if necessary drift rate), but additionally requires a search over pulse start time and DM[7]; therefore it is substantially less efficient than a search for narrowband beacons.[8]

### 2.5    Persistence.

With current telescope and computing capabilities it is impossible to reliably determine the direction of arrival when an artificial signal enters the telescope. The telescopes are indeed pointed in specific directions, yet bright sources can leak into the receiver through sidelobes of the telescope. In optical photography, this phenomenon is known as lens flare (Figure 3).

---

[6] In the numerical domain where signals are sampled regularly with a common time interval, an FFT-based poly-phase filter bank is generally used to approximate the Fourier Transform.

[7] See footnote 3.

[8] Our analysis attempts to leave out any "value judgments" placed on the relative success of a narrow band or narrow pulse search. The authors believe that until we have discovered at least one transmitting civilization, our opinions about the likelihood of one signal type being more probable than another are simply opinions.

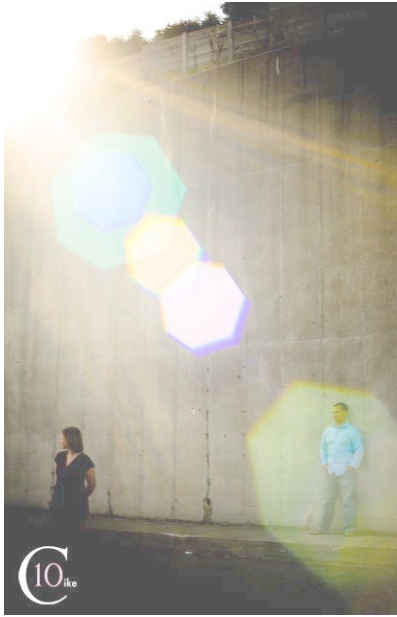

Figure 3: A photo of two people is dominated by artifacts caused by the sun. In radio astronomy imaging, similar effects occur not only from the sun but from strong satellite transmitters, and other human made transmitters in the air or on the ground. Photo Credit: Carla Ten Eyck Photography, http://photosbycarla.blogspot.com/2008/08/lens-flare-tutorial.html.

In this photo most of the field is dominated by lens flare from the Sun, a bright source at the edge or even outside of the image field. The streaks and artifacts resulting from the photo apparatus in Figure 3 have analogs in radio imaging, corrupting the radio signal measured from a single point (or pixel) on the sky.

With sufficient computational power plus many more dishes or omni-directional antennas, it might be possible to determine with high confidence the direction of arrival of a signal with only one observation. The state of the art, however, requires telescopes to re-observe the same signal over and over, with changes in pointing, focus, and then corroboration from other observatories before a signal can be identified as having extra-planetary origin. Practically speaking, signals must be persistent over days or weeks (ideally, forever). If signals are not persistent, then we cannot prove their alien origin and for safety we classify them as human-made. Given the signal fading problems described above, current radio SETI searches are limited to relatively close by, or enormously powerful transmitters. The authors hope that this limitation may one day be overcome with a clever algorithm (sooner) or by brute force (later).

Interesting signals without persistence are observed *thousands* of times each day at the SETI Institute. Figure 4 for example shows a result obtained in a narrowband SETI search near the PiHI frequency (the number $\pi$ times the HI observing line of 1420.4 MHz). This (extremely powerful) ~10 second pulse of narrowband radiation appeared in one 50 second observation period but was never re-observed. This pulse has interesting features: It is observed at a magic frequency in the direction of a nearby and potentially habitable star. Yet we cannot be sure this signal was created intentionally or unintentionally by some transmitter on Earth. Hence after multiple observations over 2 weeks and no re-detection, we gave up (although this direction is added to a catalogue of directions to re-observe as time permits).

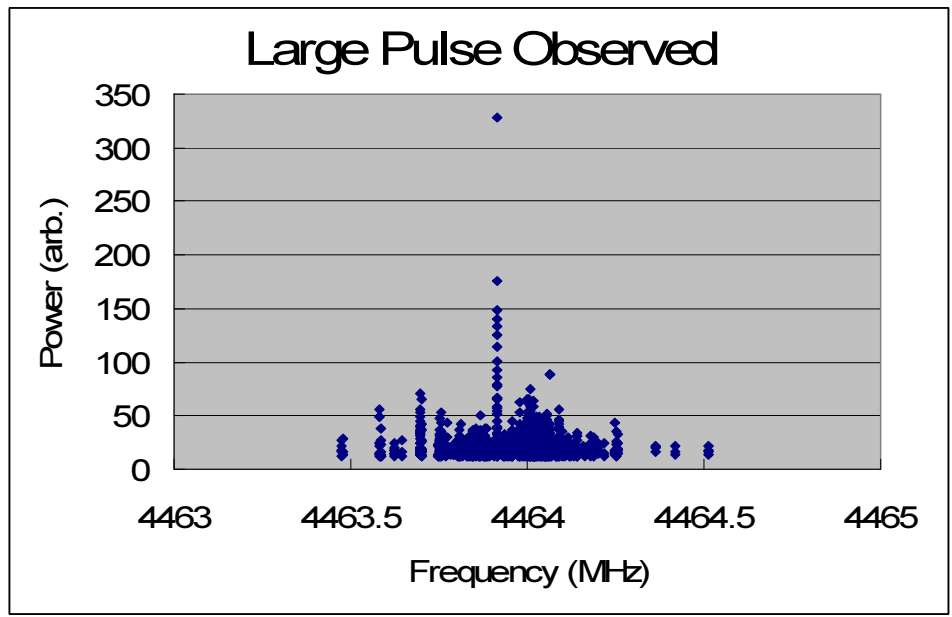

Figure 4: A pulse with maximum power >300σ above the noise background was observed on a nearby star (~100 LY, (J2000 RA, Dec) = (22.441734 hr, 79.812852$^0$)) in the HabCat Catalog (Turnbull, 2003). This pulse is interesting since it appears to arrive from the direction of a potentially habitable star and because it appears very close (within the expected Doppler shift tolerance caused by relative motion) to the "magic" PiHI frequency of 4462.3 MHz, this signal appeared in only one observation and never thereafter. Given the proximity of this source, we do not expect substantial fading in the ISM; hence the signal is really not present, most of the time.

2.6     Dimensionality of Search.

In a crude way, we can talk about computational efficiency in terms of the number of physical parameters (or dimensions) that must be searched in order to complete a survey for a subtype of ET signals. We have already discussed four physical dimensions of search for "frequency compact[9]" signals: time of arrival, emitted center frequency, Doppler drift (relative acceleration between transmitter and receiver), and DM. To this list we add two more: signal bandwidth and symbol dilation. Symbol dilation[10] is most easily understood in the context of a binary encoded signal which consists entirely of 0's and 1's. ET is free to choose the time duration of a single bit and interval between bits, from very slow (milli-Hertz or less) up to the Nyquist limit for the transmitter bandwidth.

For the most general signal type, it is necessary to search over all 6 of these dimensions, but special signal types have been discovered that "project out" some search dimensions and lowers the computation time. For example, continuous narrow band signals require only a 2-dimensional search space (frequency and Doppler drift). Extremely narrow (e.g. ps-ns) pulsed signals require a 4-dimensional search space (time of arrival, frequency, Doppler drift, DM).

---

[9] By frequency compact we mean that in whatever bandwidth is chosen for a signal, the frequency occupancy over that bandwidth is on the order of unity. We will not consider signals with very sparse frequency support in this paper.

[10] The major contributing factor to symbol dilation searches is the fact that we don't have a timing reference from the transmitting civilization. However, another important factor is dilation caused by the relative velocity of transmitter and receiver. The symbol dilation can sometimes be recovered from autocorrelation. However, the relative velocity changes with time, so not only do we have to divine the symbol rate, but its time derivative.

To help with this discussion we present a list of the number of different values each parameter must take for the 6 search parameters mentioned above as constrained by the parameters of the ATA and our current processing capabilities (1-10 GHz, observation length ~100s, narrowest frequency channel bandwidth ~1 Hz, maximum frequency bandwidth = 100 MHz) in Table 1.

Table 1: List of the relative sizes of parameter search spaces for the 6 observations variables defined above. Recall that $N_S$ is the number of samples in the measurement. However, a computation time cannot be obtained from a simple product of the numbers on the right hand side, see text for details.

| Search Parameter | Approx number of parameter values |
|---|---|
| Time of arrival | 100 s @ 10 ns sampling = $10^{11}$ |
| Center frequency | 10 GHz at 1 Hz sampling = $10^{10}$ |
| Frequency Bandwidth | Up to 100 MHz = $10^8$ |
| Dispersion Measure (DM) | Of order $N_S$ = $10^{10}$ |
| Relative Acceleration (Doppler rate) | Up to 10 Hz/s at 10 GHz = 2000 |
| Symbol dilation | Of order $N_S$ = $10^{10}$ |
| Autocorrelation Signal Duration (Repetition rate, "Gold Code," see section 3.) | 100s @ 1 µs steps = $10^8$ |

This analysis is oversimplified since not all search dimensions are comparably constrained. In a narrowband frequency search the number of different acceleration values (~200 for a 100 second measurement at 1 GHz with 1 Hz resolution) that must be probed for a reasonably complete SETI search is much less than the number of different frequencies (9 billion 1 Hz bins in the range 1-10 GHz). Yet we can say that a pulse search is always computationally less efficient than a narrowband search since pulse searching requires the same search variables as for narrowband, plus time of arrival and DM. As mentioned below, certain algorithms can take e.g. two search dimensions and reduce them to one-dimension as far as computation is concerned. This argument was first pointed out to the authors by David Messerschmitt.

3   A New Beacon Proposal.

There is a substantial historical bias toward narrowband searches in radio SETI. In the Project Cyclops report (Oliver, 1973), this seminal SETI document reads, "Beacons … will surely be highly monochromatic." The later book SETI 2020 proposed for searches including narrowband signals and high bandwidth pulsed signals. About wideband signal varieties such as frequency-shift encoding and spread spectrum SETI 2020 reads, "such a scheme would be so computationally intensive that our searching for stars and frequencies would [be very inefficient]." In this context, we now introduce a specialized set of wideband signals that do not suffer this negative consequence envisioned in SETI 2020.

A key goal of our work is to minimize the number of dimensions over which we search. Above we noted that narrowband searches are more effective than pulse searches since the latter requires searches over start time and DM while the former does not. Is there a wideband signal that can be recovered with a computational cost similar to a narrow band search?

Consider this: The transmitter sends an *arbitrary* signal with arbitrary length and arbitrary bandwidth, communicating nonzero symbol rate. After a short delay (< 1s), an adjacent transmitter (or the same one, see below) sends a second copy of the signal. The two signals are superimposed at the receiver. We assume that transmission began sometime far in the past and continues until sometime far in the future, and that the bandwidth of the transmission equal to or larger than that of our detector. We call this a twice-sent signal. Because the timescale for ISM fluctuations is on the order of seconds to hours (J. M. Cordes, Lazio, T. J. W., 1992; Walker, 1998), we can take advantage of the fact that both signals suffer *identical* dispersion and scattering by the ISM, provided the delay between transmissions is less than 1 second. We can discover such signals with an efficient autocorrelation algorithm.

The autocorrelation spectrum of a signal $A(t)$ as a function of signal delay $\tau$ is computed from

$$A(\tau) = \int_0^{N_S} S(t) \, S(t-\tau) \, dt ,  \qquad (1.3)$$

where $S(t)$ is the received signal as a function of time, and the time interval between samples is scaled to unity.[11] We assume that the detector emits a regularly sampled voltage and $N_S$ is the number of samples. $A(t)$, also known as the delay spectrum of the above described signal, will show a strong peak where $\tau = \tau_0$ the delay between the transmission of the signals mentioned above.

As an example, consider Figure 5 (left) where one message is sent twice with a delay of 7 samples. In this simulation we do not consider the effects of noise. Calculating the autocorrelation of the received signal using (1.3) gives the result on the right. On the right there is a strong peak (beacon) with power equal to half the total transmitted power of the composite signal. Using this scheme, the transmitting civilization has the opportunity to send us a great deal of information such as the complete works of Shakespeare or the complete embodiment of their society's knowledge. The message never has to repeat in order to discover this beacon; we require only that the transmitter stays on.

---

[11] For the sake of clarity, we use integral notation even though the numerical computations are performed as summations.

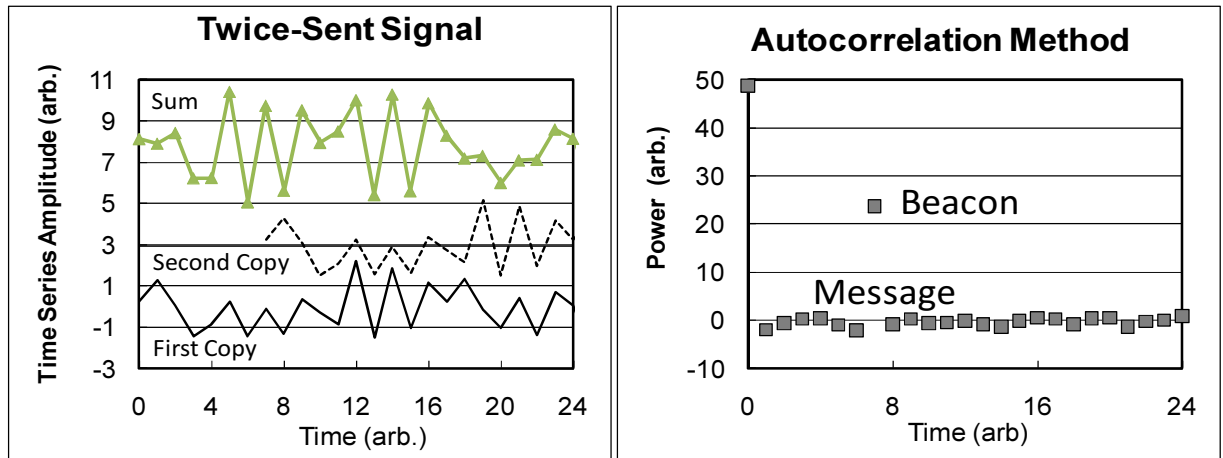

Figure 5: Left: An arbitrary signal (in this case, a sampling of Gaussian random noise) is transmitted beginning at time 0. A second copy of this signal is transmitted at time 7. These signals are superposed at the receiver. This simulation includes no galactic or receiver noise. Right: The autocorrelation spectrum computed from the superposition of the data on the right (including 4096 samples). The zero-delay autocorrelation ($\tau = 0$) is a measure of the total signal energy and the spike at ($\tau = 7$) appears when the delay is equal to $\tau_0$, the time difference between the copies. We have used Gaussian noise in this simulation to show that the method does not depend on information content in the original signal. Finally, the "noisy" background in the autocorrelation spectrum is not noise at all, it results from the accidental correlation this particular time-bounded sample. As integration time increases ($N_S$ grows larger), the beacon stands out higher and higher above this noise-like background, without limit.

More than two copies of the signal could be sent with delays of $\tau_0, 2\tau_0, 3\tau_0,$ etc., creating multiple copies of the beacon at regularly spaced delay values. Multiple-copy encoding is explicitly used for space communications so as to provide robust error correction during information retrieval. For multiply sent signals the analysis is modified to search for a series of peaks all having the same delay separation.

However, a multiply-sent signal has more redundancy hence reduced information carrying capacity[12], so there is a trade-off between redundancy and information content. For

---

[12] Here we are focusing on multiply-sent signals where all copies of the signal have the same amplitude. In this case, there are frequency ranges where the signals interfere destructively and can lead to zeros in the received frequency spectrum. If different copies are sent with different amplitudes, 90° phase offset (electric field is a real-valued function), polarization, etc. then it is

beacon discovery, there is no signal to background advantage for multiple copy signals (though there is an advantage for information retrieval due to increased redundancy). Furthermore we must guess (or search) over the number of delay copies which makes the beacon search more complex. If the number of delay copies is not 2 as in the twice sent signal, then we suggest that the next most likely value is infinity. Having an infinitude of repeats means that the beginning (preamble) of the signal is always available in any snippet of data. However, unless the time duration of the message is finite and less than the detection time, no part of the data would be recoverable since there would be fewer sampled symbols than the number of symbols representing the signal. For this reasons we favor the twice-sent signal type over any other number and we believe this paper marks the first suggestion of this approach.

Another alternative use of multiply-sent signals is an encoding scheme with amplitude-shift, frequency-shift, or phase-shift keying where a snippet of information is repeated over and over without overlap, but with regular breaks in amplitude, frequency or phase. For example, in binary phase shift keying the phase of the transmitted signal is changed (or not changed) at regular intervals to indicate the transition between two different symbols of information (a 0 or a 1). This scheme is used in GPS communications and in Section 5 we present real observations that demonstrate how autocorrelation uncovers signals of this type. Other types of encoding that use repeating signals can also be discovered using autocorrelation. An important point to our

---

possible to arrange the carrying capacity to be not reduced, but the signal to background ratio is still impacted. From the viewpoint of data extraction in a noisy environment, partial destructive interference still limits the information carrying capacity, hence the tradeoff remains.

proposal is that it is not necessary to know what code snippet is being repeated to make the discovery.[13]

Besides encoding information, twice- or multiply-sent signals have other advantages. No search over Doppler drift, time of arrival, signal dilation, or symbol rate is required. Finally, even high levels of distortion caused by the intergalactic, interstellar, interplanetary media or earth's ionosphere and troposphere can be tolerated since short-timescale correlation is highly resistant to signal distortion. This means that we could detect (but perhaps not decode) signals from much farther away than the limit for narrow band signals set by interstellar scattering mentioned in 2.3. Again, as long as both copies of the signal are distorted in the same way, they will correlate. Distortion may be a problem for signal recovery, however, as described below.

As a comparison with the narrowband and pulse searches described above, a search for autocorrelation peaks in a twice-sent signal requires the production of an autocorrelation spectrum and then a thresholding step. Since we look for strongest autocorrelation peak apart from $\tau = 0$, the size of the search space is $N_S$. As we shall see below, autocorrelation uses the same fast algorithms as for a narrowband SETI search and can be accomplished in equal time with twice as much computational power.

---

[13] In another approach, we are aware that David Messerschmitt has suggested that one *can* guess the code snippet. It may be, for example, the first 100 digits of the binary representation of the number $\pi$ or $e$. After choosing the code snippet, signal encoding proceeds as usual. We will not pursue this interesting suggestion in this paper.

### 3.1 Signal, Background Radiation, and Noise

The transmitted signal has the form $S_{\text{trans}}(t) = s(t) + s(t - \tau_0)$ where $\tau_0$ is the delay introduced between the two copies of the signal by the transmitting civilization. In a realistic example, noise is introduced by the galactic background radiation and by the receiver itself, and we detect $N_S$ samples of the received signal $S(t)$:

$$S(t) = s(t) + s(t - \tau_0) + N(t). \tag{1.4}$$

$N(t)$ is the sum of galactic background and receiver noise, and can be assumed to be Gaussian white noise to a good approximation in most cases. We define the measured variances

$$\langle s^2 \rangle = \frac{1}{N_S} \int_0^{N_S} |s(t)|^2 \, dt$$

$$\langle N^2 \rangle = \frac{1}{N_S} \int_0^{N_S} |N(t)|^2 \, dt \tag{1.5}$$

Performing the autocorrelation transform (1.3) and admitting the possibility that the computations are performed using a complex representation of the measured signal, we obtain

$$\begin{aligned}
AC(\tau) = & \int_0^{N_S} S(t) \, S^*(t - \tau) \, dt = \\
(a) \quad & \int_0^{N_S} \left[ s(t - \tau_0) \, s^*(t - \tau) \right] dt \\
(b) \quad & + \int_0^{N_S} \left[ s(t) \, s^*(t - \tau) + s(t) \, s^*(t - \tau_0 - \tau) + s(t - \tau_0) \, s^*(t - \tau_0 - \tau) \right] dt \\
(c) \quad & + \int_0^{N_S} \left[ N(t) \, s^*(t - \tau) + N(t) \, s^*(t - \tau_0 - \tau) + s(t) \, N^*(t - \tau) + s(t - \tau_0) \, N^*(t - \tau) \right] dt \\
(d) \quad & + \int_0^{N_S} N(t) \, N^*(t - \tau) \, dt.
\end{aligned} \tag{1.6}$$

The right hand side is divided into four parts, each of which contributes in its own way to $AC(\tau)$. Beginning with $(a)$, this is our "SETI signal" or beacon term, and this term reaches a maximum value of $N_S \langle s^2 \rangle$ when $\tau = \tau_0$.

For other values of delay $\tau$, $(a)$ has statistical properties similar to the three terms in $(b)$. We wish to estimate the behavior of these terms as a function of $N_S$, but this is impossible without detailed knowledge of $s(t)$, so we make some reasonable guesses. If we assume that the transmitting civilization sends us a signal "dense" with information, then it may have statistical properties similar to Gaussian white noise. As mentioned above, the signal must contain some redundancy to permit decoding, but suppose that every so often they send a preamble which gives us a "key" to decode the rest of the information. Then most of the time, all the terms $(b)$ (and term $(a)$ when $\tau \neq \tau_0$) will grow as $\sqrt{N_S} \langle s^2 \rangle$ as $N_S \to \infty$. While we cannot rely on this exact scaling behavior, we can rely on the aliens to make reasonable choices about how the information is encoded. Indeed, encoding schemes where all the terms in $(b)$ grow less slowly than $\sqrt{N_S}$ may be possible.

Before continuing, we consider a special encoding scheme where a snippet of data is sent over and over many times with a constant repeat rate. In this case, all the terms in (a) and (b) will take their turns integrating up coherently and we observe a comb of autocorrelation spikes much that first suggested by Drake (Drake, 1965). See Section 5 for a real-world example of this kind of beacon.

We identify the terms $(c)$ and $(d)$ as "noise" terms since they all contain an integral over a product where one of the terms is Gaussian white noise. For now we shall make the assumption that $s(t) \ll N(t)$ since the transmitter is far from the receiver and swamped by the galactic background. In this case, $(d) \gg (c) \gg (b)$, and we may neglect the terms in $(c)$ (and $(b)$ for twice-sent signals and the noise signal is dominated by $(d)$. Since the noise is Gaussian white distributed, the leading background noise term as $N_S \to \infty$ is $\frac{\sqrt{N_S}}{2}\langle N^2 \rangle$.[14]

With this analysis and the assumptions mentioned above we can estimate the signal to background ratio $SBR(\tau = \tau_0)$ (the ratio of the beacon peak to a typical background point in the autocorrelation spectrum) to be

$$\lim_{N_S \to \infty} SBR(\tau = \tau_0) = 2\sqrt{N_S} \frac{\langle s^2 \rangle}{\langle N^2 \rangle} \qquad (1.7)$$

To summarize, $AC$ will contain a strong peak for $\tau = \tau_0$ that is $\sqrt{N_S}$ greater than the a typical $AC$ value where $\tau \neq \tau_0$. This is our main mathematical result. In a comparison with conventional narrowband SETI, the factor of 2 on the right hand side is cancelled out by the fact that the transmitted power in $s(t)$ is only half of the total transmitted power for a twice-sent signal.

Speaking of a conventional SETI, over short time periods the scaling with $N_S$ is the same as for a "matched filter" detection algorithm. Matched filters are known to be optimal linear filters for maximizing signal to noise ratio (here, signal to background ratio). It is straightforward to perform a computation for the signal to background ratio in a narrowband search to be

$$SNR_{NB} = N_S \frac{\langle s^2 \rangle}{\langle N^2 \rangle}. \qquad (1.8)$$

---

[14] Here we use the property that $N(t)$ and $N(t - \tau)$ when $\tau \neq 0$ are independent Gaussian random numbers.

Thus the scaling law for twice-sent signals as a function of number of samples $N_S$ is not as favorable as for conventional narrowband SETI. However, in narrowband SETI the maximum integration interval is 1 second since we are open to Doppler drift rates as high as 1 Hz / second. Beyond a 1 second interval, a search over Doppler drift correction is required, which increases the complexity of the algorithm. There is no limit to the coherent integration time in the autocorrelation method.

As an aside, recent research on alternative "matched filtering" approaches to discovery of message-bearing SETI signals (Messerschmitt, 2010) show that optimal $SBR$ scaling with $N_S$ (Equation 1.8) can be obtained at the cost of substantially greater signal processing. Specifically, one must search over start time, symbol rate, Doppler drift and DM.[15] In this approach, one guesses at least part of the message content (e.g. first 100 bits of the binary representation of $\pi$) and matches to that. In this paper we cannot do justice to this developing field and leave its exposition to the future.

### 3.2 Implementation of Autocorrelation Spectroscopy.

The implementation of autocorrelation spectroscopy is a trivial extension of the processing required for conventional narrowband SETI as shown in Figure 6. In narrowband searches, the measured signal is Fourier transformed (FT) with a filter bank based on the fast Fourier transform. The resultant frequency power spectrum (PS) is formed by squaring the results of the FT and then performing a threshold operation (upward pointing arrow in Figure 6).

---

[15] David Messerschmitt has developed an algorithm where the start time and DM searches can be combined into, essentially, a 1-dimensional search. This kind of clever algorithm may possibly be expanded to further reduce the computation load in matched filter searches.

These signals that pass through the threshold detector are declared "interesting" and then followed up upon, until the direction of arrival can be confirmed.

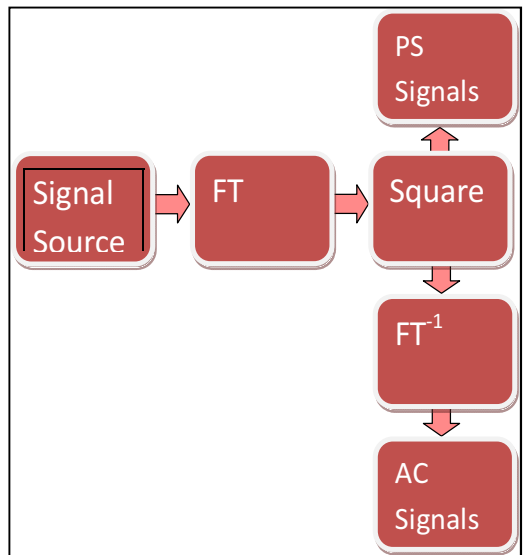

Figure 6: A block diagram showing the processing steps to recover narrowband SETI signals using the power spectrum (PS), and its relationship with the processing steps to recover twice-sent beacons using autocorrelation (AC). The figure indicates that the first three steps of processing are identical. For twice-sent signal detection, the thresholding step takes place after a second (inverse) Fourier Transform. Since the Fourier transform step is usually the most computationally intensive step, AC spectroscopy takes about twice the compute power and can be accomplished in the same time using the same computational programs as narrowband SETI.

To perform the autocorrelations required for this proposal, we take the same power spectrum values and perform an inverse FT (or $FT^{-1}$) and pass the result through a thresholding filter (or a comb filter for multiple-copy versions of the proposed signal type, see 5.1.1). A search for autocorrelated signals can be carried out simultaneously with a conventional SETI search using the same FT engine (with a trivial sign inversion) and the same thresholding detector software. We propose that in future narrowband SETI search systems, autocorrelation spectroscopy should be added for a total cost increase of less than a factor of 2.

4    Signal Recovery.

Once the beacon signal has been identified, all the arguments about search spaces fade since the transformation of a single known ET signal is negligible compared to the massive computations required to find the signal in the first place (see ).

Here we discuss some ideas for recovery of the message in a twice-sent beacon. The superimposed twice-sent beacon resulting from the signals in Figure 5, left can be described as a message (first signal) convolved with a pair of delta functions separated by $\tau_0$ in time. This convolution can be represented as a multiplication in the frequency domain (convolution theorem). In practice, we perform a discrete Fourier transform (DFT) on both the message and a second "filter" of equal length which contains only zeros except for two Dirac delta functions separated by $\tau_0$ and having the value $1/\sqrt{2}$. (We use this normalization so that the total power in the filter sums to unity.) The Fourier transforms of both signal and filter are multiplied bin by bin, and the product undergoes an inverse DFT. The resulting signal is the superposed twice-sent signal and this could be fed directly into a single transmitter for communication with Earth (i.e. two different transmitters are not necessary, though they may be convenient).

Some information is lost in this convolution. The Fourier transform of a double delta function $F_{\tau_0}(f)$ is described by

$$F_{\tau_0}(f) = z'\left[1 + \exp(i\, 2\pi\, \tau_0\, f)\right] = z \cos(2\pi\, \tau_0\, f) \tag{1.9}$$

where $z$ is a complex number with magnitude $\sqrt{2}$. $F_{\tau_0}(f)$ takes the value 0 when $\tau_0 f = \frac{1}{2}, \frac{3}{2}, \ldots, n+\frac{1}{2}$, so whatever information was present at these frequencies is destroyed by the convolution. Of course, the transmitting civilization knows this, and they are likely to use only frequency bands where the cosine function is high, perhaps in areas where $F_{\tau_0}(f) > \frac{1}{\sqrt{2}}$.

Note that these zeros could be entirely avoided by sending the second signal with different amplitude than the first. If the first signal is transmitted with amplitude 1 and the second with amplitude b, then the areas of destructive interference in frequency space would take the value $(1-|b|)^2$. Though not as dramatic, this still has negative consequences on signal recoverability in the presence of background radiation and noise. Choosing $b \neq 1$ also reduces the beacon $SNB$ from its maximum value at $b=1$.

In this section, we have not yet discussed the impact of background noise on message recovery. The effects of noise are draconian. We first define the Fourier conjugates of the message $s(t)$ and noise $N(t)$ to be $s(f)$ and $N(f)$, respectively. Unless $s(f) > N(f)$ for a given frequency, it is impossible to recover the message information for that frequency. However, if we have identified a SETI beacon using autocorrelation, then not only do we know $\tau_0$ but we know we are dealing with a SETI beacon. If necessary, humanity will build a sufficiently sensitive radio telescope to achieve $s(f) \gg N(f)$. This might not be necessary – recall that an autocorrelation search at radio frequencies has yet to be performed (or at least published). To promote the discussion we simply assume $s(f) \gg N(f)$ from here on.

We define the signal to background ratio in the frequency domain $SBR(f)$ is[16]

$$SBR(f) = \frac{\langle s(f) \rangle}{\langle N(f) \rangle} \cos(2\pi\tau_0 f) \qquad (1.10)$$

---

[16] Here we take $SNB$ to be the quantity which is used during the thresholding process, where we look for peaks that stand out far above the background level. This threshold is set by choosing a specific value of $SNB$. To calculate the probability of false alarm rates, a different quantity (approximately the square of $SNB$) may be required.

Since we know positions of the zeros of $s(f)$ we can estimate $\langle N(f) \rangle$. Using this information and the peak positions of $s(f)$ we can estimate $\langle s(f) \rangle$. Now we define a dimensionless threshold $P$ (~10) which determines the reliability we wish to achieve in the message recovery. We obtain poor estimates of the message anywhere $SBR(f) < P$, so we must discard this data (e.g. set to zero or something similar). The result $s'(f)$ is then divided by $F_{\tau_0}(f)$ to remove the distortion caused by the double delta function. Finally, an estimate of the original message $s'(t)$ is obtained from an inverse Fourier transform of $s'(f)$. Alternatively, the civilization may choose to encode their message in the frequency domain, where our message estimate is $s'(f)$. In either case, we may have a very good estimate of the message, provided that the transmitting civilization has encoded their message to avoid the zeros of $s(f)$. Notice that for any $N < \infty$ there will be finite probability of an error in the signal recovery. This defect might be further mitigated if the transmitting civilization chooses a message with inherent redundancy that allows error correction in signal recovery.

There is more that could be said about the fascinating topic of message recovery and how to optimize it. Here we present only the simplest possible approach as a demonstration of the feasibility and limits of message recovery for the twice-sent beacon.

5    Proofs of Principle Using the Allen Telescope Array.

As follow through on this proposal as a search for interstellar beacons we have begun observations of known point-source emitters such as pulsars, masers, galaxies and human made satellites. For now we choose directions where there are known strong point source emitters, with the idea that some of the received radiation may have embedded autocorrelation information that is not apparent in ordinary radio astronomical, narrow band or narrow pulsed

SETI observations. Hence we have hope of detecting autocorrelation power even in a relatively short measurement of duration ms to minutes.

For an astronomical example we show a measurement of methanol maser emission in the W3OH molecular cloud region in Figure 7. On the left we show the part of the frequency power spectrum where methanol maser emission is found. On the right we show the autocorrelation spectrum. The autocorrelation contains bumps and oscillations associated with the rather narrow 100 kHz frequency bandwidth of the maser signal. In Figure 7 we plot time delays of up to 150 μs since this is the most interesting region. However, the transmitting civilization must use a value of $\tau_0$ much greater than the delay range of the maser spectrum so that it will stand out from the background. We did not discover any interesting features for larger values of delay, up to fractions of a second in these prototype experiments. In future reports we will show results from longer time integrations and more sources.

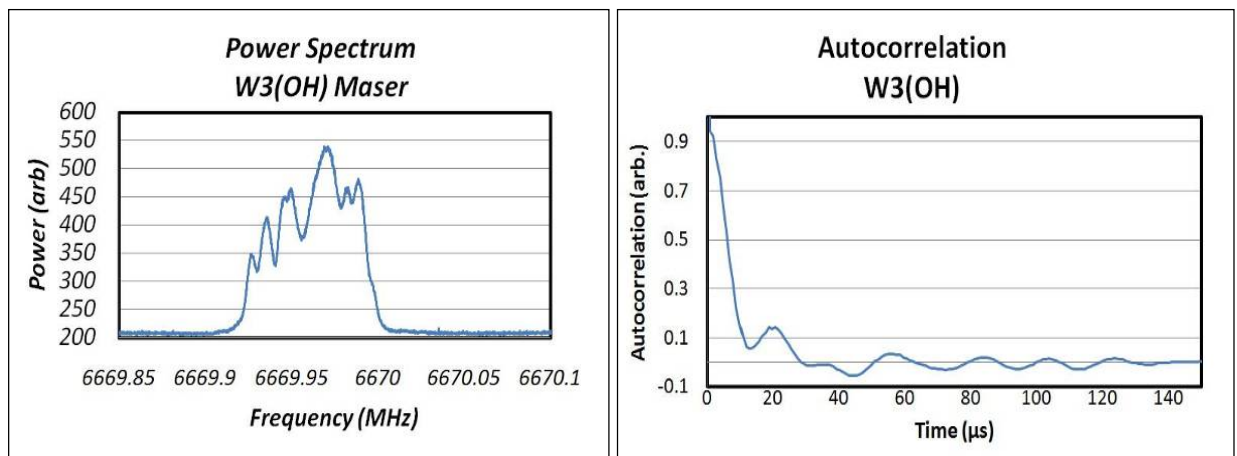

Figure 7: Frequency power spectrum (left) and autocorrelation spectrum (right) of the molecular cloud W3OH in the frequency range of methanol maser emission. In the power spectrum, we observe evidence of multiple maser clouds with different temperatures (line widths) and different relative velocities (Doppler frequency offset). In the autocorrelation spectrum, we see artifacts of the relatively narrow frequency support of the signal on the left.

The concept behind measuring masers is that an extraterrestrial civilization might "pump" such masers to cause amplitude variations in the maser power. Tommy Gold proposed this idea (article not available) in late 60's and early 70's, and this idea was shown to be feasible by Joel Weisburg (Weisberg, 2005) who found a maser being modulated by pumping from a nearby pulsar.

In a second example, we display in Figure 8 the power and autocorrelation spectra from a GPS satellite. GPS communication uses binary phase shift key encoding where each bit of information is represented by 20 copies of a 1023-bit "gold code" sent sequentially with a repeat rate of 1 ms. Although the artificial nature of this signal is hardly apparent in the power spectrum (and over longer time periods would be even smoother), the artificial nature of the received signal is obvious in the autocorrelation spectrum on the right. This is precisely the kind of AC spectrum expected for multiple-time-sent SETI signals and indicates intelligent origin, in this case human origin.

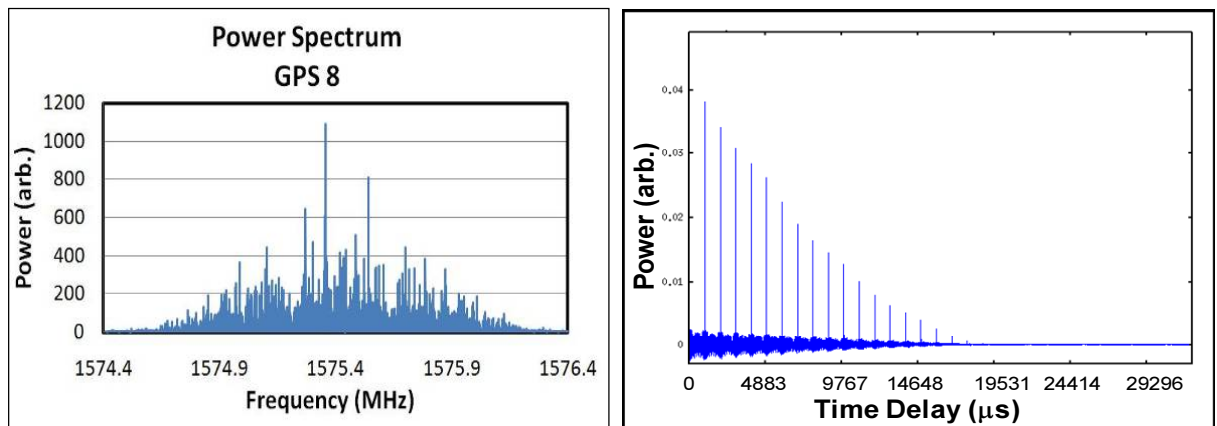

Figure 8: Power spectrum and autocorrelation spectrum from a GPS satellite known as PRN08. The broad frequency support on the left looks fairly noise-like within an envelope of ~2 MHz. Hence a narrowband SETI search would be confounded by this artificial signal. However, the autocorrelation spectrum shows a series of peaks separated by 1 ms which is clear evidence of this signal's artificial origin.

### 5.1.1 Generalizations.

We have already discussed many generalizations of the twice-sent beacon, including multiply-sent versions and repeating code schemes (e.g. binary phase shift keying). One can consider giving the twice-sent signals different amplitudes. This would decrease the detectability of the beacon, but would allow, at least in principle, full recovery of all message data as a function of frequency or time. More complex schemes can be constructed with multiply-sent signals copies having different amplitudes, polarization[17], prime number delay ratios etc. Such approaches may decrease detectability and/or maximum message symbol rate, but may improve signal recovery.

In another direction, twice-sent beacons could be redundant in the frequency domain rather than in the time domain, just as narrowband signals are the frequency-domain equivalent of single pulses. However, redundancy in the frequency domain implies that the two signal copies will not be identical upon arrival due to differential ISM-related scattering (and the delay will be changed by dispersion, which could cause trouble for signal recovery). We expect even better ideas based on our simple suggestion will be forthcoming both from our group and other groups performing research in this field.

6    Conclusions.

We propose a new class of beacon signals that contain rich information content while standing out substantially from the galactic background radiation. As beacons, they are highly resistant to distortions during the voyage from transmitter to receiver. These signals are detectable with a simple and efficient autocorrelation algorithm. Although autocorrelation techniques for signal detection have been speculated upon in the past, here we provide <u>observational</u> evidence with a GPS satellite that demonstrate the feasibility for detecting such beacons and identifying them as having intelligent origin.

---

[17] For polarization, it is best to choose circularly polarized polarizations since the ISM plasma is also influenced, in general, by an unknown magnetic field along the direction of travel. This introduces (another) delay between the two signal copies.

We describe an implementation of an autocorrelation detector that can run simultaneously on the same data stream in a narrowband SETI search and uses the same computational blocks. Thus both narrowband SETI and autocorrelation SETI can be run together in real time with less than a factor of 2 augmentation of compute resources. Finally we present an introduction to the challenges faced by humans once these signals are detected. Solutions and suggestions about how to analyze data for signal recovery are discussed. Because these signals contain substantial information, they are a more straightforward method to actually communicate (one way) useful information from the extra-solar civilization to human kind.

Conventional narrowband SETI and pulse searches are promising ways to look for extra-solar civilizations. At the same time, alternative searches like the one proposed here are worth investigation and merit the relatively small additional effort to carry them out in parallel.

Acknowledgements: The first phase of the ATA was funded through generous grants from the Paul G. Allen Family Foundation. UC Berkeley, the SETI Institute, the National Science Foundation (Grant No. 0540599), Sun Microsystems, Xilinx, Nathan Myhrvold, Greg Papadopoulos, and other corporations and individual donors contributed additional funding. This work was also supported by the NSF through award AST-083826 and by NASA award NNG05GM93G. The authors gratefully acknowledge David Messerschmitt and Ian Morrison for fruitful conversations and constructive criticisms of the poster associated with this work.

7    References.

Backer, D. C., Kulkarni S. R., Heiles C., Davis M. M., Goss W. M. (1982). *Nature, 300*, 615.

Backus, P. R. (2010). *SETI Survey at Arecibo, Project Phoenix*. Ap.J. in preparation.

Cohen, R. J., Downs, G., Emmerson, R., Grimm, M., Gulkis, S., Stevens, G., and Tarter, J.C. (1987). Narrow Polarized Components in the OH 1612 MHz Maser Emission from Supergiant OH-IR Sources. *Royal Astronomical Society, Monthly Notices, 225*, 491-498.

Cordes, J. M., Lazio, T. J. W. (1992). NE2001.I. A New Model for the Galactic Distribution of Free Electrons and its Fluctuations, 2010, from http://arxiv.org/abs/astro-ph/0207156

Cordes, J. M., Lazio, T. J. W., & Sagan, C. . (1997). *ApJ, 487*, 782.

Drake, F. D. (Ed.). (1965). *Current Aspects of Exobiology: The Radio Search for Intelligent Extraterrestrial Life*: Pergaman Press.

Fitzpatrick, R. (2006). Classical Electromagnetism: An intermediate level course, 2009, from http://farside.ph.utexas.edu/teaching/em/lectures/node100.html

Jackson, J. D. (1975). *Classical Electrodynamics*. New York: John Wiley & Sons.

Lazio, T. J. W., Tarter, J. C. and Backus, P. R. (2002). Megachannel Extraterrestrial Assay Candidates: No Transmissions from Intrinsically Steady Sources. *Astronomical Journal, 124*, 560-564.

Messerschmitt, D. (2010), unpublished.

Narayan, R. (1986). Maximum Entropy Image Restoration in Astronomy. *Ann. Rev. Astron. Astrophys., 24*, 127-130.

Nyquist, H. (1928). Certain topics in telegraph transmission theory. *Trans. AIEE, 47*, 617-644.

Oliver, B. M., Billingham, J. (1973). *Project Cyclops: Design Study of a System for Detecting Extraterrestrial Life* (Vol. NASA Technical Report CR-114445): NASA.

Shannon, C. E. (1949). Communication in the presence of noise. *Proc. Institute of Radio Engineers, 37*(1), 10-21.

Siemion, A., Von Korff, J., McMahon, P., Korpela, E., Werthimer, D. , Anderson, D., ,Bower, G., Cobb, J., Foster, G., Lebofsky, M., van Leeuwen, J., and Wagner, M. (2010). Acta Astronautica.

Thompson, A. R., Moran, J. M., Swenson, G. W. (2001). *Interferometry and synthesis in radio astronomy*. New York: Wiley-Interscience.

Turnbull, M. a. T., J.C. (2003). TARGET SELECTION FOR SETI. I. A CATALOG OF NEARBY HABITABLE STELLAR SYSTEMS. *ApJ Suppl., 181-198*.

Walker, M. A. (1998). Interstellar scintillation of compact extragalactic radio sources. *Mon. Not. R. Astron. Soc., 204*, 307-311.

Weisberg, J. M., Johnston, S., Koribalski, B., and Stanimirovic, S. (2005). Discovery of Pulsed OH Maser Emission Stimulated by a Pulsar. *Science, 309*, 106-110.